\documentclass[manuscript]{acmart}
\makeatletter
\def\@ACM@copyright@check@cc{}
\makeatother
\setcopyright{cc}

\usepackage{array}
\usepackage{tabularx}
\usepackage{ltxtable}
\usepackage{graphicx}
\usepackage{booktabs}
\usepackage{makecell} 
\usepackage{stfloats} 
\usepackage{xspace} 
\usepackage{subcaption}

\copyrightyear{2026}
\acmYear{2026}
\setcopyright{cc}
\setcctype{by}
\acmConference[CHI '26]{Workshop paper submitted to 'Herding CATs: Making Sense of Creative Activity Traces' at the 2026 CHI Conference on Human Factors in Computing Systems}{April 13--17, 2026}{Barcelona, Spain}
\acmBooktitle{Herding CATs: Making Sense of Creative Activity Traces Workshop of the 2026 CHI Conference on Human Factors in Computing Systems (CHI '26), April 13--17, 2026, Barcelona, Spain}
\acmDOI{}
\acmISBN{}

\begin{document}

\title{From State Changes to Creative Decisions: Documenting and Interpreting Traces Across Creative Domains}

\author{Xiaohan Peng}
\orcid{0000-0002-5202-419X}
\affiliation{%
  \institution{LISN\\ Université Paris-Saclay, CNRS, Inria
  }
  \city{Orsay}
  \country{France}
}
\email{xiaohan.peng@inria.fr}

\author{Sotiris Piliouras}

\orcid{0000-0001-6918-7836}
\affiliation{%
  \institution{LISN\\ Université Paris-Saclay, CNRS, Inria
  }
  \city{Orsay}
  \country{France}
    }
\email{sotirios.piliouras@inria.fr}

\author{Carl Abou Saada Nujaim}
\orcid{0009-0007-0558-6118}
\affiliation{%
  \institution{LISN \\
  Université Paris-Saclay, CNRS, Inria}
  \city{Paris}
  \country{France}
  }
  \email{carl@lisn.fr}

\begin{abstract}
Analyzing creative activity traces requires capturing activity at appropriate granularity and interpreting it in ways that reflect the structure of creative practice.
However, existing approaches record state changes without preserving the intent or relationships that define higher-level creative moves. 
This decoupling manifests differently across domains: GenAI tools lose non-linear exploration structure, visualization authoring obscures representational intent, and programmatic environments flatten interaction boundaries. 
We present three complementary approaches: a node-based interface for stateful GenAI artifact management, a vocabulary of visual cues as higher-level creative moves in visualization authoring, and a programming model that embeds semantic histories directly into interaction state.
\end{abstract}




\maketitle

\section{Introduction}
Analyzing creative activity traces requires capturing activity at appropriate levels of granularity and interpreting them in ways that reflect the structure of creative practice. 
Existing approaches such as version histories, interaction logs, or collaborative records tend to record state changes without preserving the higher-level intent or semantic context that give those changes meaning. 
Traces thus remain structurally decoupled from the creative decisions they result from across creative domains and tool paradigms: losing non-linear exploration structure in GenAI tools, obscuring representational intent in visualization authoring, and reducing meaningful actions to low-level operations in programming environments.

In this paper, we present a multi-faceted approach to documenting and interpreting creative traces in GenAI, visualization, and programming contexts.
First, we introduce a node-based interface design that treats generative artifacts as stateful, manipulable units, enabling designers to preserve alternative versions, branch explorations, and link semantic attributes to visual content. 
Second, we examine how 3D authoring traces such as material appearance, perspective, deformation, and dynamics can be mapped to higher-level ``creative moves'' that reflect representation and framing choices in expressive 3D visualizations.
Finally, we propose a programming model that embeds semantic histories directly into interaction state, supporting both single-user and collaborative environments.

\section{Documenting GenAI Creative Traces: Tracking Interactive Generative Visual Artifacts across Scales}
While prior creative history management systems introduce git-style version control, semantic grouping~\cite{AdaptiveHistoryChen16}, and node-based workflows~\cite{angert2023:spellburst}, they are primarily oriented toward capturing discrete, linear iterations and state transitions as the system evolves~\cite{DesignManagerYou25}, or often focus on relatively simple graphic elements~\cite{rawn2023:Quickpose} rather than higher-abstraction, context-dependent and hierarchically structured design assets.
We argue for more visual, dynamically applicable and retrievable, and non-linear histories tailored to iterative, GenAI-supported design, particularly as GenAI images become a new class of design assets.
\begin{figure}[h!]
    \centering
    \includegraphics[width=0.7\linewidth]{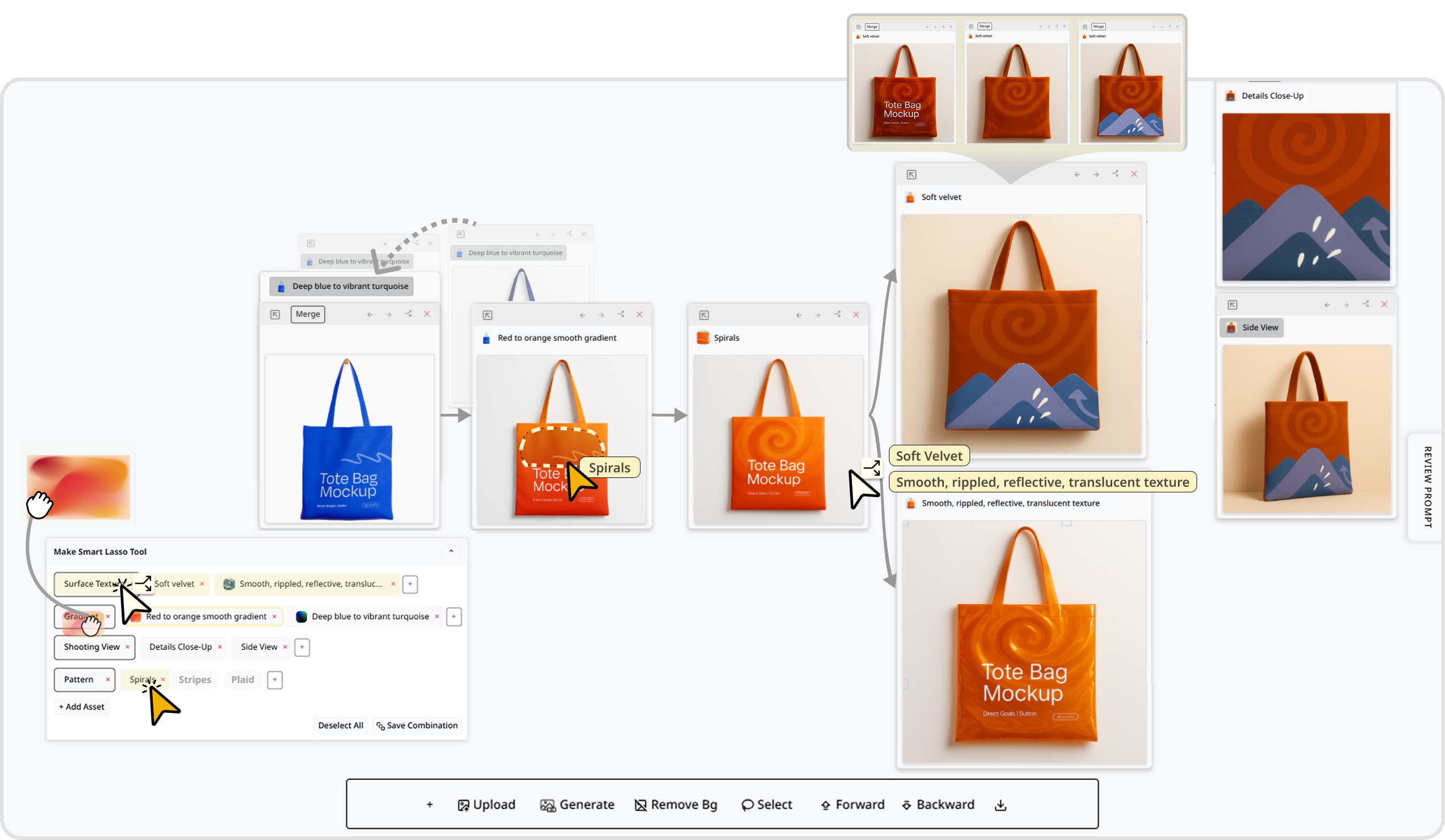}
    \caption{DesignTrace: a GenAI design tool for exploring alternatives and tracking design progress. Designers can extract or create semantic attributes then apply them to visuals, branch multiple alternatives under shared semantic categories, make localized edits inside individual canvas node while navigating editing history and different versions. 
    }
    \label{fig:DesignTrace}
    \Description{}
\end{figure}
\subsection{Stateful Nodes for Dynamic Visual Interaction and Documentation}
Existing node-based systems typically treat nodes as parametric operators in a data-flow pipeline. However, in \textit{DeisngTrace}~\cite{peng2026:designtrace}, we explored nodes as interactive and stateful artifacts placed on a freeform canvas, each containing an editable image canvas and a local history of generation and editing actions. 
They can be independently resized, collapsed, duplicated, and spatially arranged to externalize design states as discrete, manipulable units. 
This structure enables designers to keep alternatives visible, compare them side by side, branch manually or intelligently with semantics, and reuse prior states without clutter or overlap. 
By linking semantic attributes to visual content and preserving branching histories, Nodes support non-linear exploration while documenting the evolving rationale behind design decisions.
\subsection{Multi-Scale GenAI Traces Combine Commands, Artifact States, and Exploration Paths}
Current chat-based GenAI tools' logs can grow long easily since they weigh minor edits and major design shifts equally, while canvas-based ones can get cluttered without proper design on the prompt-image relationship.
\textit{DesignTrace} supports multi-scale trace management. 
Each node maintains a local history of edits, while persistent, reusable nodes represent intermediate states that can be spatially arranged and revisited.
By connecting these nodes into single or multiple exploration paths, the system provides an overview of design progression while preserving flexibility for branching and reuse.
This structure enables designers to trace minor temporal-proximal changes, but also trace ``design versions,'' while keeping a broader overview of different exploration paths.

\section{Visualizing Creative Traces: Representation and Framing Choices in Expressive 3D Visualization Authoring}
3D creators iterate through intertwined decisions about geometry, materials, lighting, camera/viewpoint, and motion expressive in 3D visualization authoring.
Yet creative activity traces are typically captured as sequences of parametric edits but not the higher-level representational intent behind those creative choices.
Seemingly small edits can correspond to high-level representation and framing choices with substantial novelty, such as scattering objects in 3D space to communicate accumulation of microplastics in the ocean~\cite{scicomlab_explore_ocean} or using particle motion simulations to convey air pollution~\cite{youtube_dtqsi_plgxoa}.

Reconstructing creative intent from parametric edits is challenging because most 3D software is representation-agnostic, and logs lack the semantic and representational structure needed to relate actions to visualization goals, especially for unconventional representations.
We propose that interpreting creativity traces in expressive 3D visualization can benefit from an intermediate vocabulary that translates tool actions into representation-level ``creative moves''. 
Our ongoing work on physically-inspired visualization~\cite{piliouras2025incorporating} provides a toolkit for this translation by describing (i) the visual elements that compose a scene, (ii) the physically-inspired properties being manipulated, and (iii) their semantic relationship to the data variables and theme (such as literal, iconic, or symbolic relationships). 
In this workshop, we want to explore how this structured design space of material-based visual representation choices can be used to make sense of creative activity traces in expressive 3D visualization authoring, supporting the analysis of 3D exploration trajectories and informing future creativity-support tools.

\section{Programming Creative Traces: Embedding Semantic Histories in Interaction State}
Digital creative activity traces---ranging from changing the color of a shape to generating a new design alternative in a GenAI tool---remain difficult to analyze because they are typically captured as external logs rather than produced by the system's interaction model.
Even in collaborative environments based on Conflict-Free Replicated Data Types~\cite{shapiro_conflict-free_2011} (CRDTs), histories record operational changes but not the structure of user activity, requiring developers to manually reconstruct intent by grouping operations into higher-level transactions.
Inspired by Object-Oriented Drawing~\cite{Xia2016:ObjectOrientedDrawing}, we introduce a model for embedding histories directly in application state and take a first step toward representing semantics in both single-user and collaborative environments.
\begin{figure}[h!]
    \centering
    \includegraphics[width=0.5\linewidth]{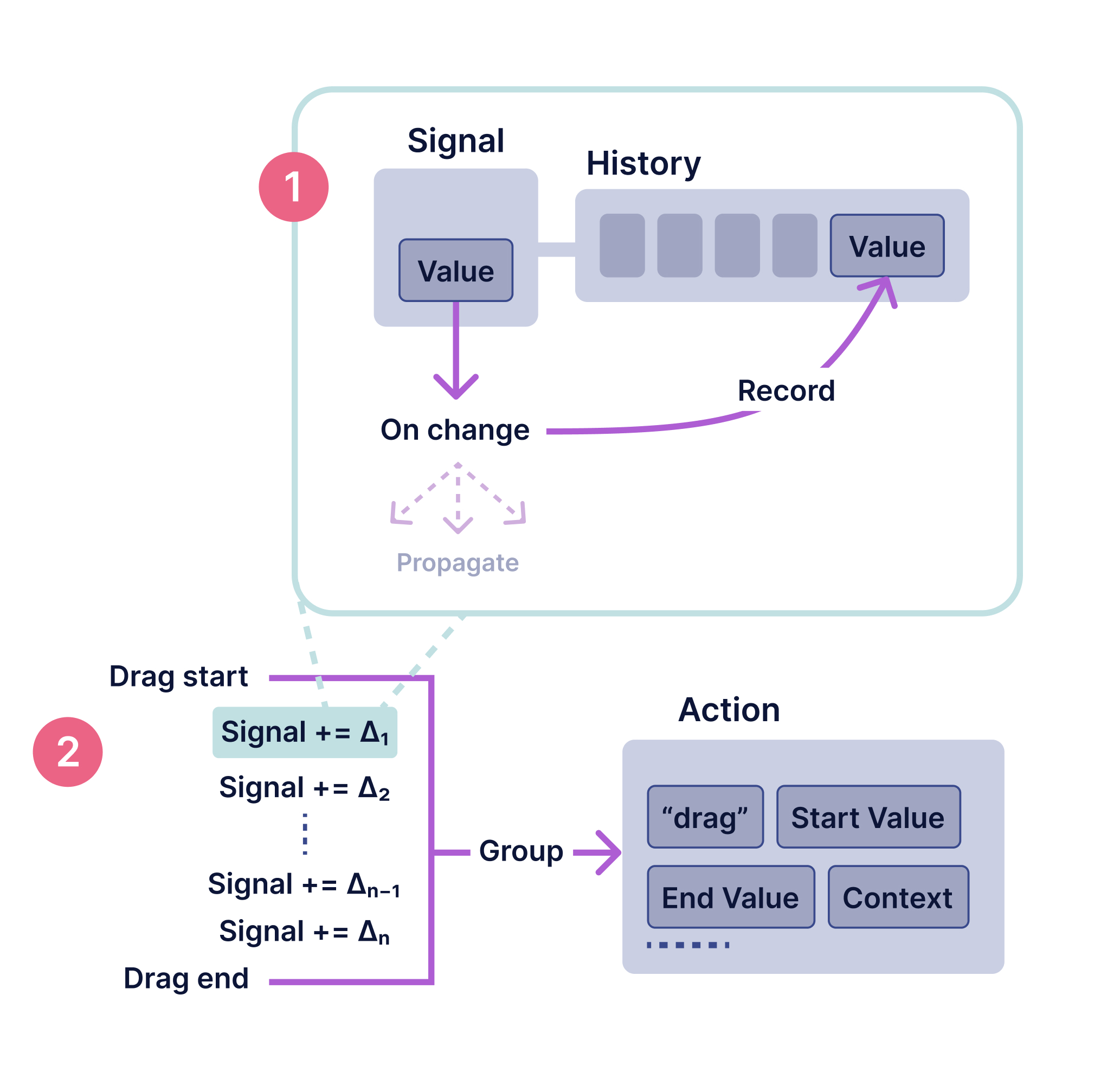}
    \caption{(1) A reactive signal records value changes in a persistent history. 
    (2) A semantic action groups a segment of history changes recorded between the start and end of an interaction, such as a drag.}
    \label{fig:ProgTraces}
    \Description{}
\end{figure}

Modern front-end frameworks increasingly rely on reactive signals~\cite{Oeyen_2024_Reactive}, primitives that propagate state changes to dependent computations, such as functions that re-render part of the interface.
As signals typically expose only the current value of a variable, we augment them with persistent histories (figure \ref{fig:ProgTraces}.1) that record all past values together with relevant metadata (e.g. timestamps).
Developers can therefore visualize, query, and replay prior states without implementing custom instrumentation.
We further group state changes according to interaction boundaries (e.g., from drag start to drag end), producing explicit action blocks (figure \ref{fig:ProgTraces}.2) that provide contextual information about user activity.
This constitutes a first step toward semantic preservation by reconstructing meaningful user actions from state evolution.
In CRDT systems such as Yjs\footnote{\url{https://yjs.dev}}, all value changes are reliably shared between collaborators and can be explicitly grouped into transactions, with undo/redo functionalities operating over these updates.
We build on this mechanism by recording a semantic action as described above along with each transaction.
As a result, the shared history contains not only what changed, but also what action was performed, allowing collaborators to follow meaningful steps instead of a stream of low-level edits.

\section{Conclusion}
This paper examined the challenge of capturing creative activity traces that reflect the structure of creative practice rather than mere state changes. 
Across three domains, we showed how this structural decoupling manifests differently and requires distinct approaches: 
stateful nodes that preserve non-linear exploration and semantic branching in GenAI design; a vocabulary of 3D design cues that surfaces representational intent in expressive visualization authoring; and a programming model that embeds semantic action boundaries directly into interaction state. 
These contributions together suggest that meaningful trace capture is less a logging problem than a design problem---one that requires rethinking how tools expose the decisions underlying creative activity.

\bibliographystyle{formalities/ACM-Reference-Format}
\bibliography{main.bib}

\end{document}